\documentclass[]{spie}  

\usepackage{amsmath,amsfonts,amssymb}
\usepackage{comment}
\usepackage{graphicx}
\usepackage{subcaption}
\usepackage[colorlinks=true, allcolors=blue]{hyperref}
\usepackage{textgreek}
\usepackage[margin=1in]{geometry}
\usepackage{booktabs}
\usepackage{array}

\title{Initial modeling and testbed feasibility studies of WaveDriver: a laser guide star AO system for HWO}

\author[a]{Dominic F. Sanchez}
\author[a]{Benjamin L. Gerard}
\author[a]{Peter Waswa}
\author[a]{Alex Geringer-Smith}
\author[b]{Aditya Sengupta}
\author[a]{Alexx Perloff}
\author[a]{Michael Messerly}
\author[a]{William Moore}
\author[a]{Matthew Cook}
\author[a]{Paul Pax}
\author[a]{Cesar Laguna}
\author[b]{Mathew DeMartino}
\author[b]{Kevin Bundy}
\author[b]{Rebecca Jensen-Clem}
\author[a]{Aaron Lemmer}
\author[a]{S. Mark Ammons}
\author[a]{Megan Eckart}
\author[a]{Lisa Poyneer}
\author[a]{Eric Strang}

\affil[a]{Lawrence Livermore National Laboratory, United States}
\affil[b]{University of California Santa Cruz, United States}

\authorinfo{Further author information:\\Dominic F. Sanchez: E-mail: sanchez107@llnl.gov}

\begin{document} 
\maketitle

\begin{abstract}
The Habitable Worlds Observatory (HWO) will require \textgreater100× greater wavefront stability than achieved by JWST, pushing conventional telescope architectures to picometer-level performance limits. WaveDriver is a mission concept that relaxes these constraints by employing an external laser guide star (LGS) spacecraft flying in formation with HWO to enable wavefront sensing and control. Early results from the High Contrast Testbed demonstrate feasibility of required stability using coronagraphs, multiple deformable mirrors, and a suite of wavefront sensors. We report mission design analyses, initial testbed stability metrics, compare sensor performance, and present an optical model that quantifies LGS brightness, conjugation errors, and field errors arising from finite beacon separation.
\end{abstract}


\section{INTRODUCTION}
\label{sec:intro}  

Over the past few decades, advances in high-contrast imaging have brought the detection and characterization of nearby potentially habitable exoplanets within reach. Detecting these planets remains challenging because they are up to $10^{10}$ times fainter than their host stars at small angular separations. While modern coronagraphs can suppress much of the stellar light, achieving the required contrast ultimately depends on extreme wavefront control and optical stability.

The next generation of ground-based observatories will provide unprecedented angular resolution and will use advanced adaptive optics (AO) systems to correct atmospheric turbulence. However, residual low-order wavefront errors produce stellar leakage that can obscure faint planet signals, particularly at the smallest angular separations. Although ground-based observatories will significantly advance exoplanet science, atmospheric turbulence fundamentally limits the achievable stability and contrast required for detecting Earth-like planets.

Space-based observatories avoid atmospheric turbulence and therefore provide the stable environment necessary for achieving the extreme contrast required for habitable exoplanet imaging. Motivated by the goals outlined in the 2020 Decadal Survey, the Habitable Worlds Observatory (HWO) has been proposed as NASA's next flagship mission dedicated to the detection and characterization of nearby habitable exoplanets. HWO will most likely have a segmented primary mirror design, and a high-performance coronagraph to achieve $10^{-10}$ contrast levels. Reaching this performance requires unprecedented wavefront stability, with temporal wavefront errors maintained below approximately 10~pm over 10-minute intervals ($\sigma_{10}$). While segment actuators can compensate for low-temporal-frequency disturbances, mid- and high-frequency wavefront errors remain a significant technical challenge \cite{douglasLaserGuideStar2019, potierAdaptiveOpticsPerformance2022}

AO provides a promising path toward actively suppressing these higher-frequency disturbances. Modern AO systems routinely operate at bandwidths 100-1000~Hz, providing temporal correction beyond the capabilities of segmented mirror control alone. Incorporating a high-speed AO system into HWO offers a potential risk-mitigation strategy by extending wavefront control into temporal regimes that would otherwise remain uncorrected. However, no natural guide stars are sufficiently bright to sense picometer-level wavefront disturbances at the required temporal bandwidths.

WaveDriver addresses this challenge by proposing a laser guide star (LGS) architecture in which one or more formation-flying spacecraft illuminate HWO with a laser beacon for high-speed wavefront sensing \cite{gerardWaveDriverLaserGuide2026}. This approach enables continuous sensing of temporal disturbances while using mature AO technologies to extend wavefront correction toward the picometer regime. In this paper, we present the WaveDriver mission concept, optical modeling of the LGS architecture, and initial laboratory measurements that demonstrate progress toward picometer-level AO for future flagship space observatories.

\section{WaveDriver Mission Analysis and Design}
\label{sec:mission-design}

\subsection{Mission Overview}
The WaveDriver mission will provide LGS illumination to NASA's HWO, enabling the $\sigma_{10}$ wavefront stability required to detect and characterize Earth-like exoplanets around nearby stars, thereby significantly expanding the known population of potentially habitable worlds. 

Instead of relying on the science star for guiding, an LGS can be used to provide increased wavefront sensing flux. This approach to wavefront sensing uses a bright calibration source as an artificial guide star, enabling high-cadence segment and/or deformable mirror (DM) control during coronagraph observations of stellar systems, independent of the host star’s brightness.

\begin{figure*}[ht]
\centering
\includegraphics[width=0.8\textwidth]{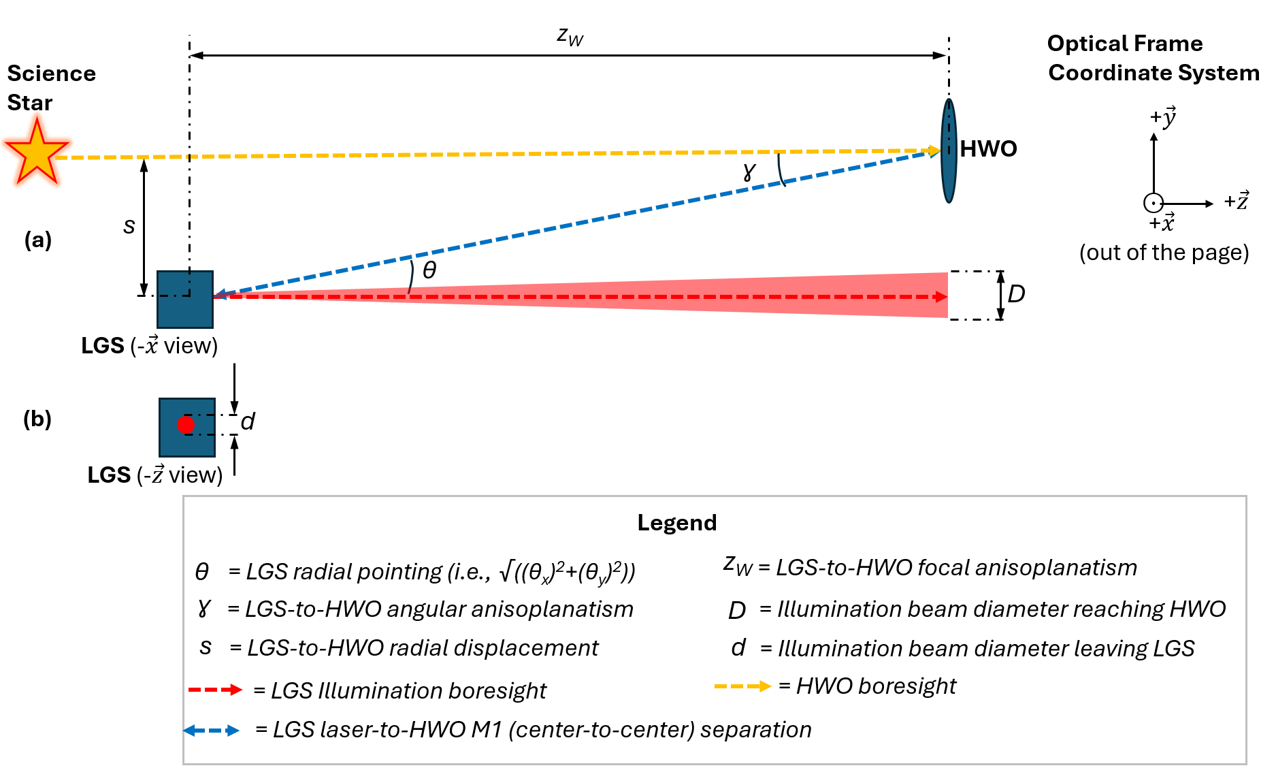}
\caption{WaveDriver mission observation geometry.}
\label{fig:observation_geometry}
\end{figure*}

Figure~\ref{fig:observation_geometry} describes the observation geometry showing angular and focal anisoplanatism parameters in the optical reference frame. The configuration depicts the relative positions of the science target, HWO telescope, and LGS spacecraft during a nominal observation session.

\subsection{WaveDriver Mission Formulation}
The following mission goals and objectives have been developed to shepherd formulation of the WaveDriver mission.
\paragraph{Mission Goal:}
Support HWO to achieve the required picometer-level wavefront stability necessary for detecting and characterizing Earth-like exoplanets around nearby stars. 
\paragraph{Primary Mission Objective:}
Illuminate the Adaptive Optics system aboard the Habitable Worlds Observatory (HWO) during science observation sessions to support HWO achieve the operational contrast stability of σ10 or better required for detecting and characterizing habitable exoplanets. 
\paragraph{Secondary Mission Objective:}
Technology demonstration—to demonstrate basic operational capability and provide proof-of-concept data for future missions, even if full operational support to HWO is limited.

\subsubsection{Requirements Development}
\begin{figure*}[ht]
\centering
\includegraphics[width=0.8\textwidth]{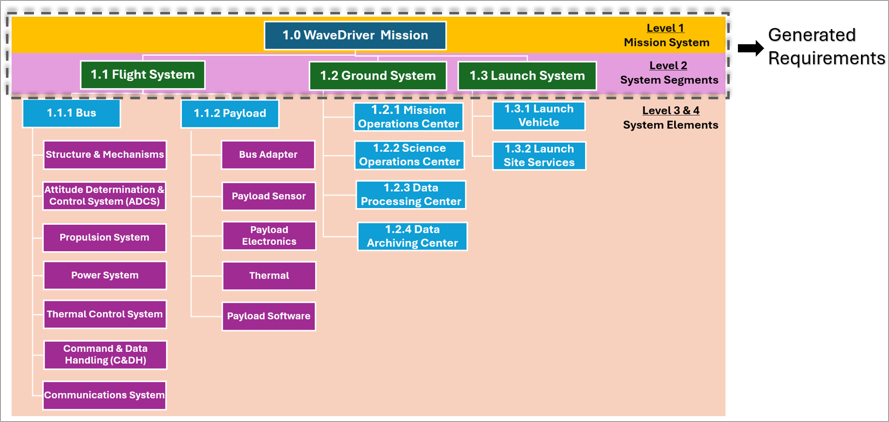}
\caption{WaveDriver requirements flow-down status.}
\label{fig:requirements}
\end{figure*}

Subsequently, to accomplish the stated mission objectives,  a hierarchical requirements framework is adopted as shown in Figure~\ref{fig:requirements}. This framework demonstrates the process of decomposing high-level requirements into more detailed lower-level requirements. The requirements flow-down approach translates the WaveDriver mission objectives into implementable requirements, organized into four hierarchical levels. Presently, eleven Level 1 requirements and 19  Level 2 requirements have been formulated and elucidated in the WaveDriver Mission Requirements Document (MRD) artifact.

The description of the four requirements hierarchical levels is given in Table \ref{Tab:requirements}.

\newcolumntype{P}[1]{>{\raggedright\arraybackslash}p{#1}}

\begin{table}[h]
\centering
\caption{Description of requirements hierarchical levels.}
\renewcommand{\arraystretch}{1.5}
\begin{tabular}{@{}P{4cm} P{9cm}@{}}
\toprule
\textbf{Requirements Level} & \textbf{Description} \\
\midrule
Level 0 - Mission Objectives & High-level science objectives defining the mission's purpose and goals [See Section 2.2] \\
\midrule
Level 1 - Mission Requirements & Mission-level scientific and programmatic requirements that fulfill the Mission Objectives. Encompasses both baseline and threshold requirements \\
\midrule
Level 2 - System Requirements & System segment requirements that define architecture, functionality, and interfaces of constituent elements required to satisfy Level 1 requirements. \\
\midrule
Level 3 \& 4 - Subsystem Requirements & Detailed subsystem and assembly specifications that satisfy Level 2 requirements, including operational conditions, interface definitions and performance parameters. \\
\bottomrule
\end{tabular}
\label{Tab:requirements}
\end{table}

\subsubsection{Mission Analysis and Design}
The mission analysis and design effort primarily focuses on conducting a mission architecture trade study. The outcome of this study will be a down-selected mission architecture(s) concept that would be further matured into a viable mission. In addition to the mission architecture definition, the selected concept entails development of the spacecraft configurations, concept-of-operations, mission costs etc.


\begin{table}[h]
\centering
\caption{Baseline HWO Mission Trajectory (parameters represent a probable HWO mission orbit profile)}
\renewcommand{\arraystretch}{1.5}
\begin{tabular}{@{}c P{4.5cm} P{7.5cm}@{}}
\toprule
\textbf{\#} & \textbf{Parameter} & \textbf{Value} \\
\midrule
1 & Orbit type & Quasi-Halo orbit at Sun--Earth L2 point \\
\midrule
2 & Earth--spacecraft distance & $\sim$1.28 $\times$ 10$^{6}$ km (EME2000) \\
\midrule
3 & Duration &
\begin{tabular}[t]{@{}l@{}}
$\sim$5.4 Years (Sept 2026 -- Mar 2032) \\[0.3em]
\end{tabular} \\
\midrule
4 & ``Orbital Period'' & $\sim$330--395 days (11--13 months) \\
\midrule
5 & HWO spacecraft velocity & 0.1--0.5 km/s \\
\midrule
6 & Reference frame & EME2000 (ECI frame) \\
\bottomrule
\end{tabular}
\label{Tab:HWO_orbit}
\end{table}

To facilitate Wavedriver mission design and  analysis it is essential to define baseline conditions that mimic the actual HWO mission as much as possible. Subsequently for our trade study analysis, the adopted baseline HWO mission trajectory and science target suite are shown in Table \ref{Tab:HWO_orbit} and Table \ref{Tab:Science_targets} respectively.

\begin{table}[h]
\centering
\caption{Baseline Science Target Suite (obtained from \citenum{Tuchow_2025} and private communications with Chris Stark).}
\renewcommand{\arraystretch}{1.5}
\begin{tabular}{@{}c P{5cm} P{8cm}@{}}
\toprule
\textbf{\#} & \textbf{Parameter} & \textbf{Value} \\
\midrule
1 & Number of targets & 210 Stars \\
\midrule
2 & Number of observation sessions & 1{,}313 Sessions \\
\midrule
3 & Science observation period & $\sim$55.7-month (7 Dec 2026 -- 31 Jul 2031) \\
\midrule
4 & Observation duration statistics &
\begin{tabular}[t]{@{}l@{}}
\textbullet\ Minimum: 0.129160 hrs. \\
\textbullet\ Maximum: 373.958400 hrs. \\
\textbullet\ Average: 21.349698 hrs.
\end{tabular} \\
\midrule
5 & Target observation Re-visit interval &
\begin{tabular}[t]{@{}l@{}}
\textbullet\ Minimum: 1234.093117 hrs. \\
\textbullet\ Maximum: 18044.410224 hrs. \\
\textbullet\ Average: 4584.190540 hrs.
\end{tabular} \\
\midrule
6 & Target Observation strategy &
\begin{tabular}[t]{@{}l@{}}
\textbullet\ Phase 1 (Detection) \\
\textbullet\ Phase 2 (Characterization)
\end{tabular} \\
\bottomrule
\end{tabular}
\label{Tab:Science_targets}
\end{table}

\begin{figure*}[ht]
\centering
\includegraphics[width=0.99\textwidth]{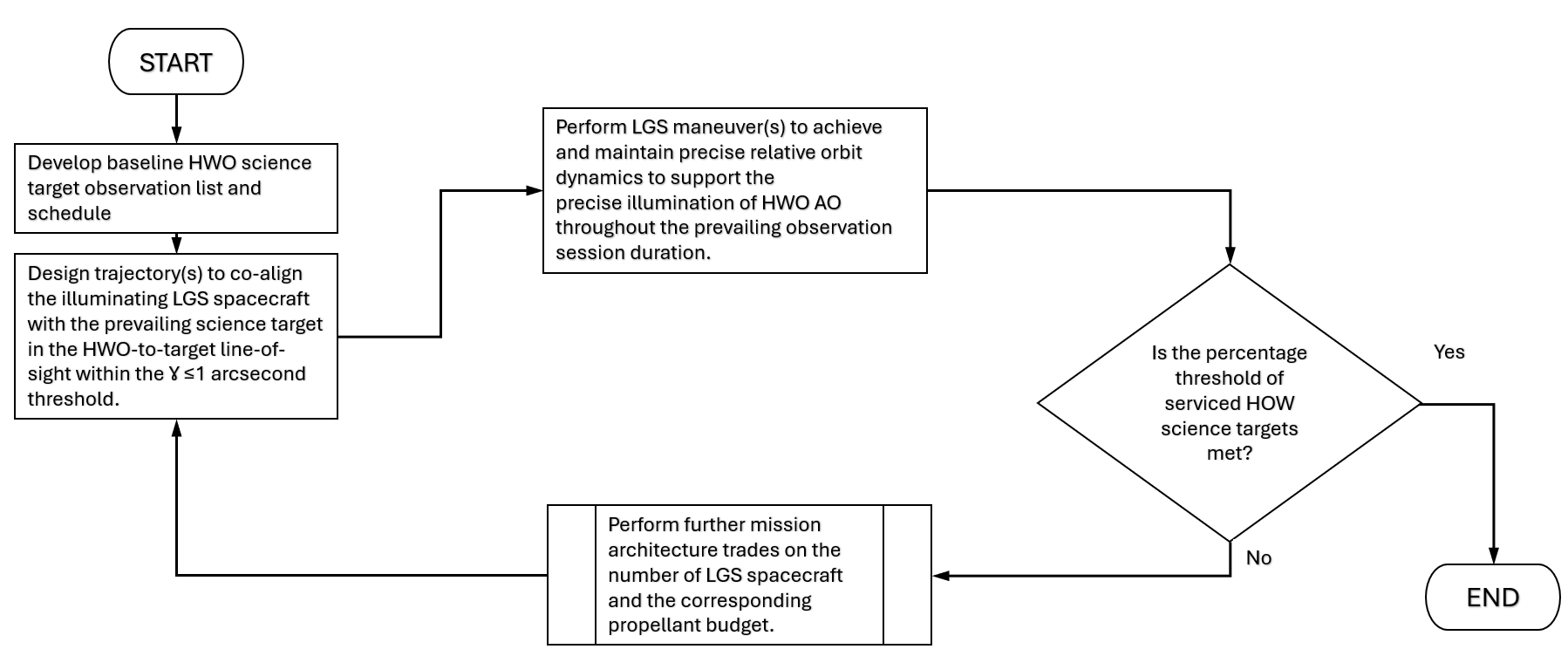}
\caption{WaveDriver mission architecture trade study methodology .}
\label{fig:trade_methodology}
\end{figure*}

The approach summarized in Figure~\ref{fig:trade_methodology} is used to conduct the mission architecture trade study. This trade study aims to develop a viable mission architecture consisting of one or more LGS spacecraft that will be able to service HWO during coronagraphic observations of exoplanet.  

\subsection{Preliminary result examples}
\begin{figure*}[ht]
\centering
\includegraphics[width=0.8\textwidth]{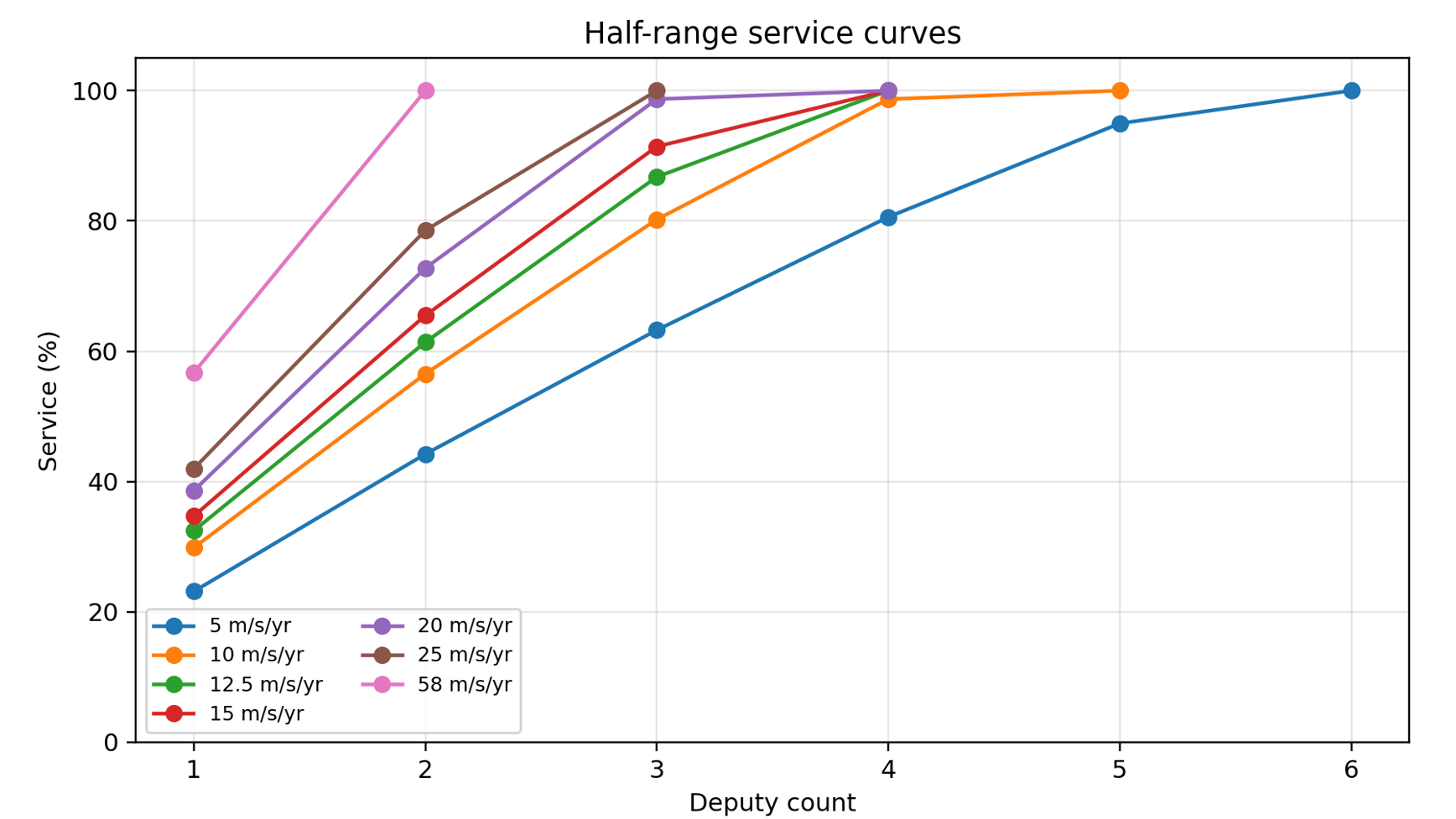}
\caption{Percentage of service accomplished as number of LGS (deputy) spacecraft increases.}
\label{fig:result_example1}
\end{figure*}

\begin{figure*}[ht]
\centering
\includegraphics[width=0.8\textwidth]{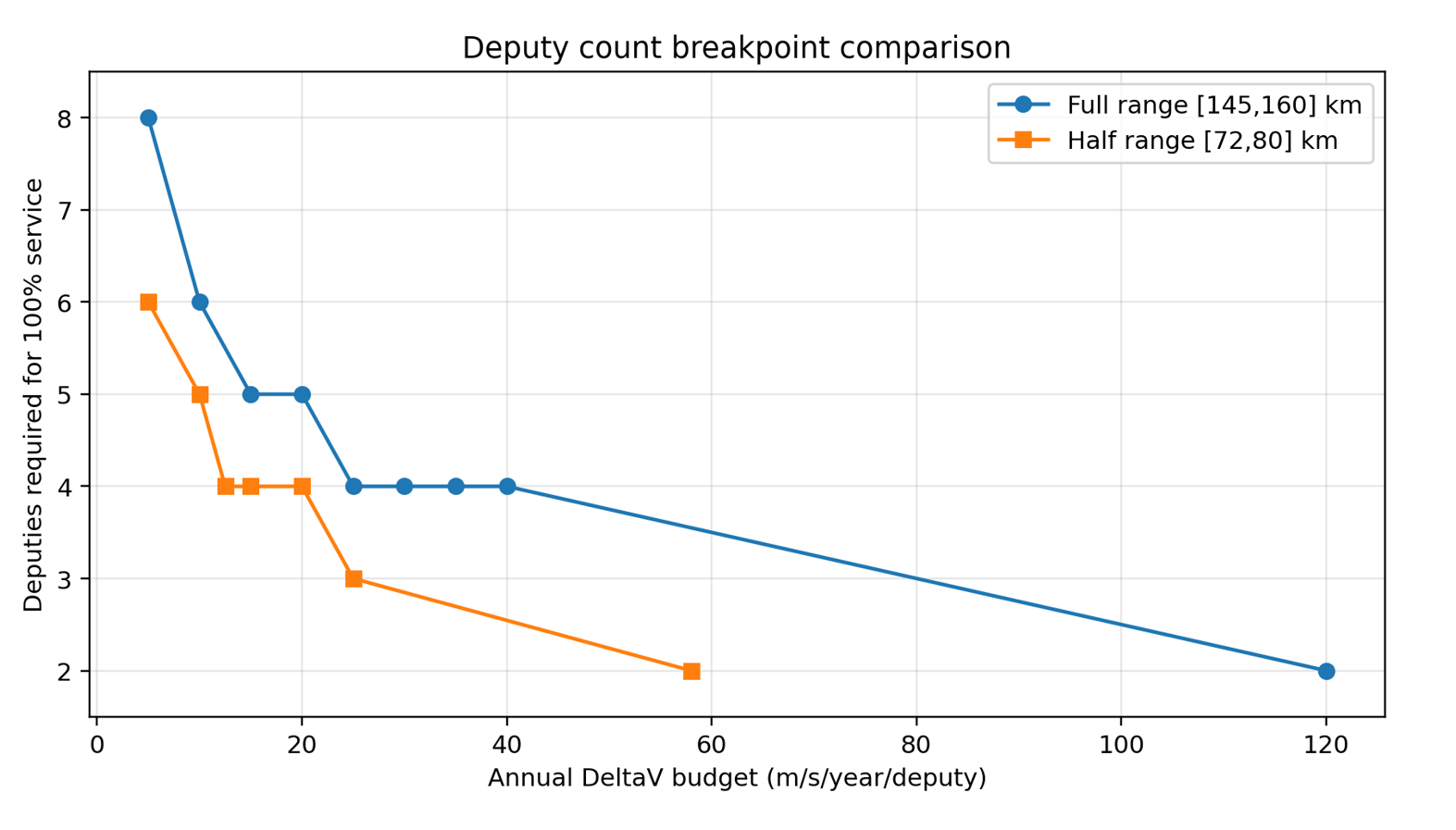}
\caption{Number of LGS spacecraft required for 100$\%$ HWO servicing as a function of annual $\Delta V$ budget.}
\label{fig:result_example2}
\end{figure*}

Figures~\ref{fig:result_example1} and \ref{fig:result_example2} are two illustrative examples of the outputs from the ongoing mission architecture trade studies.

\subsection{Highlights and Prospects}
\begin{itemize}
    \item Maintain precise relative orbit dynamics with HWO  in the unstable Sun-Earth L2 region.
    \item Maintain precise LGS wavefront stability over extended durations ($>1$ hour).
    \item Multi-spacecraft collaborative operations at Sun-Earth L2.
    \item Execute precisely coordinated observation sequences with HWO on demand without a priori  planning.
\end{itemize}


\subsection{Laser Beacon Brightness}

A fundamental requirement of the WaveDriver architecture is that the LGS provides sufficient photon flux for high-speed wavefront sensing. \citenum{douglasLaserGuideStar2019} showed that an apparent V-band magnitude of approximately -3 enables picometer-level AO by providing adequate signal-to-noise ratio (SNR) for temporal wavefront sensing. The laser spacecraft must therefore produce a beacon of sufficient brightness while maintaining a reasonable power budget.

To evaluate this trade, the apparent magnitude of a Gaussian laser beacon was calculated as a function of spacecraft separation and beam waist for a 1~mW laser. The initial beam waist determines the beam divergence, with larger beam waists producing smaller divergence angles and maintaining a higher irradiance over longer propagation distances. Conversely, smaller beam waists diverge more rapidly, distributing the laser power over a larger area at the telescope aperture and reducing the apparent brightness of the beacon.

\begin{figure*}[ht!]
\centering
\includegraphics[width=0.8\textwidth]{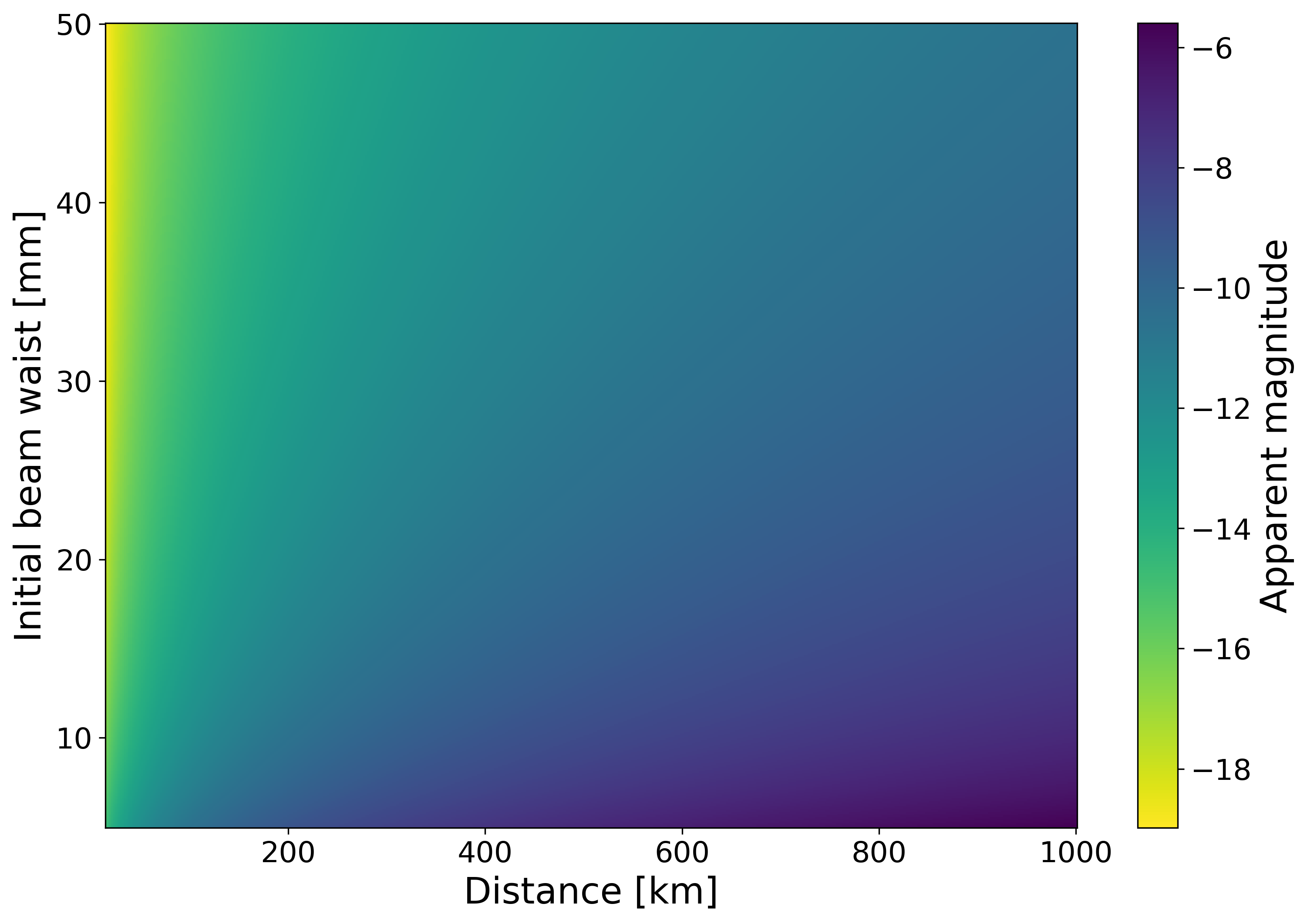}
\caption{Apparent magnitude of a 1~mW Gaussian laser guide star as a function of spacecraft separation and transmitted beam waist. Larger beam waists reduce beam divergence, maintaining a higher irradiance over long propagation distances and producing a brighter LGS. Across the explored parameter space, a modest-power laser produces a LGS brighter than the -3 magnitude requirement for picometer-level AO.}
\label{fig:lgs-magnitude}
\end{figure*}

Figure~\ref{fig:lgs-magnitude} shows the predicted apparent magnitude as a function of spacecraft separation and initial beam waist. As expected, the beacon becomes fainter with increasing spacecraft separation due to beam divergence. Increasing the transmitted beam waist partially compensates for this effect by reducing the divergence and preserving the irradiance at the telescope aperture. Over the range of beam waists from approximately 5-50 mm and spacecraft separations extending to 1000 km, a 1~mW laser remains  brighter than the -3 magnitude requirement.

These results indicate that only modest laser powers are required to produce a sufficiently bright LGS for AO, providing flexibility in the design of the laser spacecraft. The spacecraft separation and beam waist can therefore be selected primarily based on optical and mission constraints.

\section{Modeling}
\label{sec:modeling}

\subsection{Finite Conjugate Mismatch}
Because the LGS originates from a finite conjugate rather than infinity, the beam footprint expands as it propagates through the telescope. At the secondary mirror, the illuminated pupil becomes larger than that produced by an science source, causing the wavefront sensor (WFS) to sample regions of the optics that are not illuminated by the science beam. This finite-conjugate pupil mismatch, analogous to focal anisoplanatism in ground-based LGS systems, introduces non-common optical paths that lead to wavefront estimation errors and cannot be corrected by the AO system.

\begin{figure*}[ht]
\centering
\includegraphics[width=0.8\textwidth]{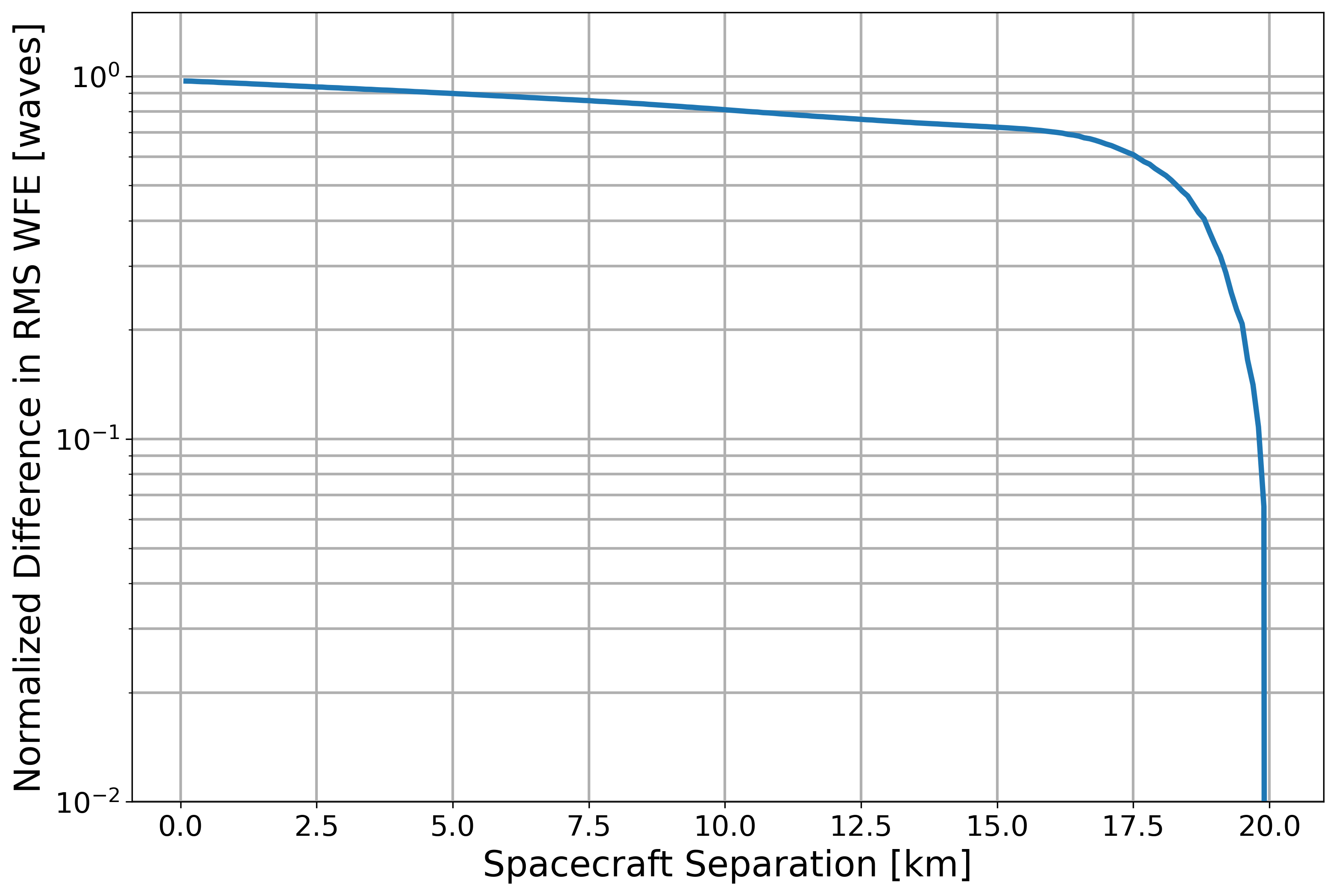}
\caption{Normalized RMS wavefront error introduced by finite-conjugate pupil mismatch as a function of LGS spacecraft separation for the HWO EAC1 optical prescription. Increasing the spacecraft separation reduces the difference between the finite-conjugate and infinite-conjugate beam footprints, which decreases the resulting non-common-path wavefront error. Spacecraft separations of approximately 15–20 km reduce the finite-conjugate error to a negligible level for picometer-level wavefront sensing.}
\label{fig:spacecracft-separation}
\end{figure*}

Figure~\ref{fig:spacecracft-separation} illustrates the relationship between spacecraft separation and the relative pupil size at the secondary mirror. As the spacecraft distance increases, the beam footprint asymptotically approaches that of an object at infinity. To quantify the resulting error, a 100~pm RMS wavefront was propagated through the optical system assuming an ideal WFS. 
Requiring the residual error introduced by the finite-conjugate geometry to remain below 1~pm implies that the meta-pupil diameter must differ from the infinite-conjugate case by less than approximately 1\%. For the EAC1 optical prescription (\citenum{Carrier2025} and obtained CODE V design through private communication with J. Tesch), this corresponds to a minimum spacecraft separation of approximately 15~km. At larger separations, the finite-conjugate beam footprint mismatch becomes negligible compared with the overall wavefront error budget.

\subsection{Field Angle Constraints}
\label{sec:title}

In addition to finite-conjugate pupil mismatch, the LGS spacecraft illuminates HWO at a field angle relative to the science target. This off-axis target causes the beam footprint to shift across the telescope optics and can introduce field-dependent aberrations. Before considering these aberrations, however, the first-order geometric constraint is determining whether the beam remains fully transmitted through the optical system without vignetting.

\begin{figure}[ht]
    \centering
    \begin{subfigure}[b]{0.45\textwidth}
        \includegraphics[width=\textwidth]{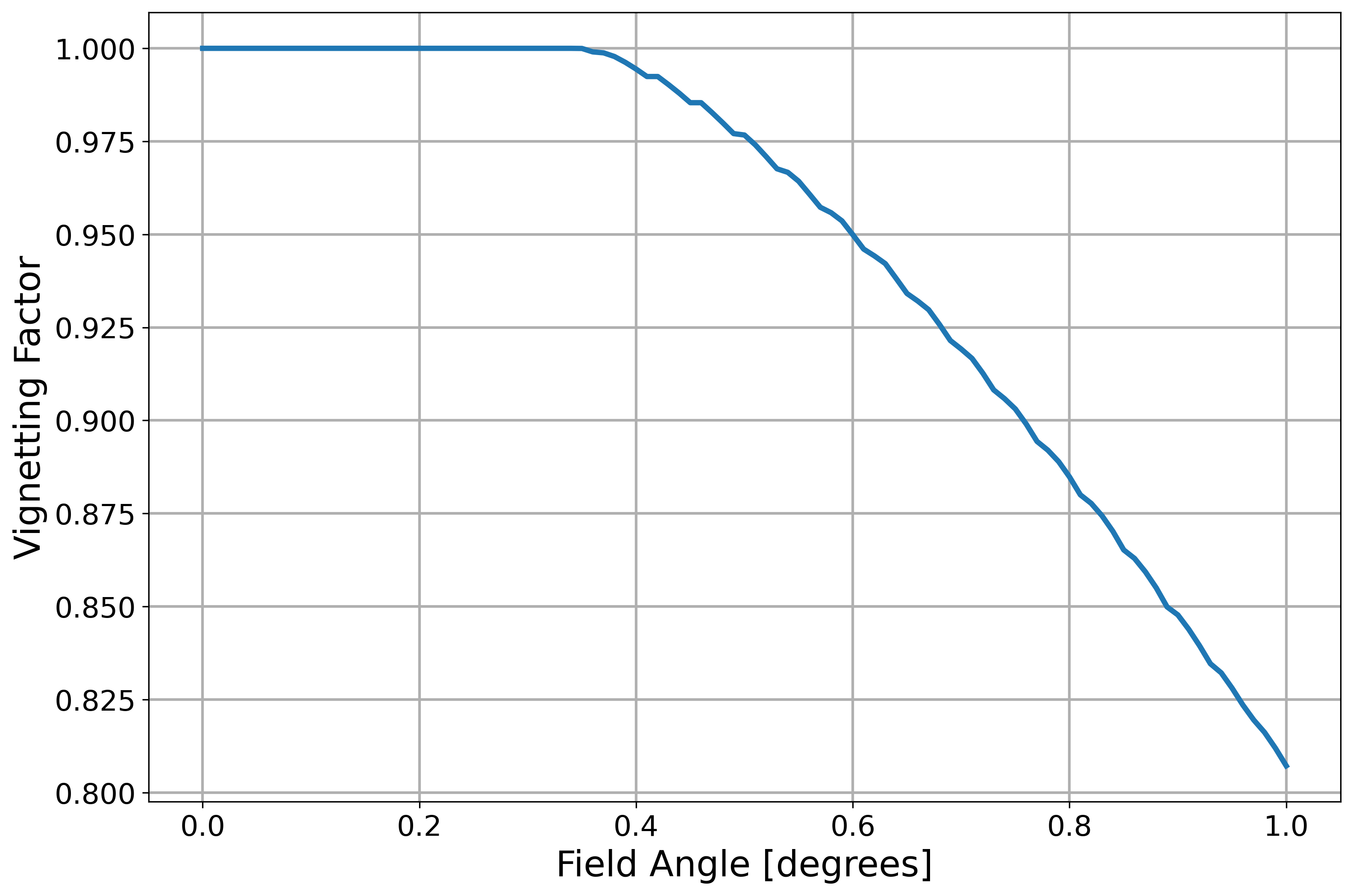}
        \caption{}
        \label{fig:vignetting}
    \end{subfigure}
    \hfill
    \begin{subfigure}[b]{0.45\textwidth}
        \includegraphics[width=\textwidth]{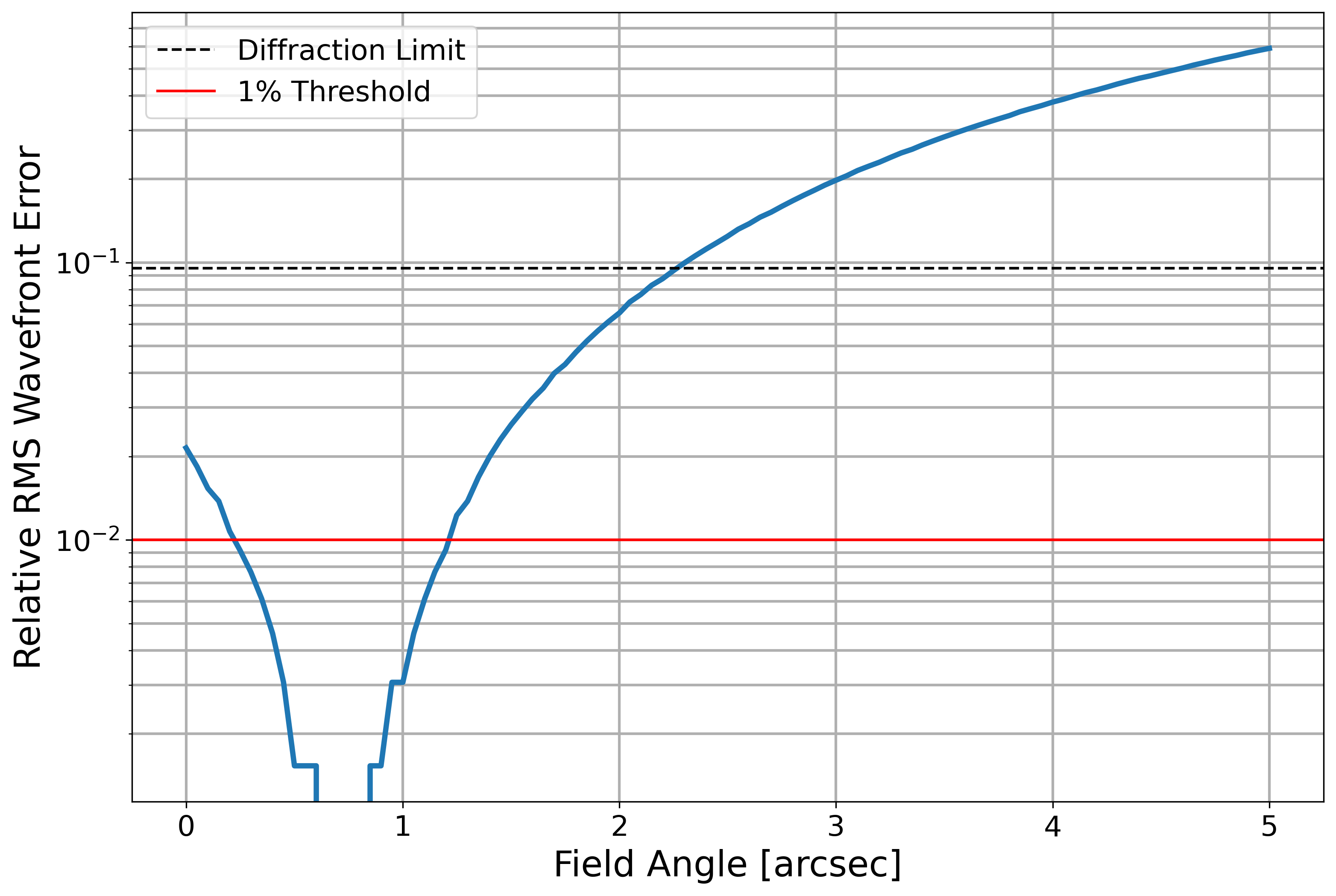}
        \caption{}
        \label{fig:rms-vs-field}
    \end{subfigure}
    \caption{Field-angle constraints for the WaveDriver LGS. (Left) Fraction of unvignetted rays as a function of field angle, showing that vignetting begins near 0.3 degrees but remains below approximately 20\% at 1~degree. (Right) Relative RMS wavefront error as a function of field angle. Field-dependent aberrations increase with angular LGS offset and become larger than the on-axis beam beyond approximately 1.2~arcsec, providing a first-order constraint on the allowable guide star pointing geometry.}
\end{figure}

Figure~\ref{fig:vignetting} shows the fraction of unvignetted rays as a function of field angle for the EAC1 telescope. In this initial analysis, only the secondary mirror was considered to illustrate this effect. Vignetting begins near 0.3°, although more than 80\% of the beam remains unvignetted even at a 1~degree field angle. Since the proposed WaveDriver architecture operates at substantially smaller angular separations, vignetting is not expected to limit the baseline mission concept. Instead, this analysis establishes an initial constraint on the allowable spacecraft pointing geometry while demonstrating that the required field angles remain well within the optical design.

In addition to vignetting, operating the LGS at a field angle introduces field-dependent aberrations that differ from an on-axis target. These non-common aberrations reduce the accuracy of the reconstructed wavefront and, if sufficiently large, limit the ability of the AO system to accurately compensate temporal disturbances. Figure~\ref{fig:rms-vs-field} shows the normalized RMS wavefront error as a function of field angle for the baseline EAC1 optical prescription, evaluated at the first focal plane. As expected, the wavefront error increases with field angle as the beam travels through different regions of the optical system. Although the quantitative relationship depends on the telescope prescription, this analysis provides an initial estimate of the allowable LGS field angle. To minimize non-common path errors, the LGS should remain within the region where the relative difference between the on- and off-axis wavefront error remains below 1\%, corresponding to field angles of approximately 1.2~arcsec for the EAC1 design.

\section{Laboratory Testbed Stability}
\label{sec:lab-demo}

\subsection{Experimental Objective}
A key challenge in maturing AO for HWO is demonstrating picometer-level wavefront correction under realistic temporal disturbances. AO can relax the telescope's passive stability requirement to approximately 100~pm and suppress these errors down to the 10~pm level. Achieving this capability requires a DM with picometer-level resolution and WFSs capable of accurately measuring temporal disturbances at these amplitudes.

The optical analyses presented in the previous section establish the geometric constraints required for a LGS AO architecture. The remaining challenge is demonstrating that the AO system itself can sense and correct wavefront disturbances at the picometer level. As an initial demonstration toward this goal, we characterize the temporal stability of the High Contrast Testbed (HCT) at Lawrence Livermore National Laboratory (LLNL). Rather than immediately attempting closed-loop correction, we first establish the fundamental stability of the testbed by measuring the temporal power spectral density (PSD) of tip and tilt, which typically dominate low-order wavefront error.

Tip and tilt were selected as the initial focus because they are often the dominant low-order aberrations affecting imaging and coronagraph performance. Demonstrating accurate measurement of these modes therefore provides a practical first milestone toward picometer-level AO while establishing the analysis framework that can later be extended to higher-order aberrations. 
 
\subsection{High Contrast Testbed}

The HCT is a fully reflective, in-air coronagraphic testbed designed for the development and validation of high-contrast imaging technologies. The testbed includes a suite of WFSs, including a Shack–Hartmann wavefront sensor (SHWFS), a three-sided reflective pyramid wavefront sensor (3RPWFS), a Zernike wavefront sensor (ZWFS), a self-coherent camera (SCC), and a pupil-chopping wavefront sensor (PC-WFS) \cite{sanchez2020,baudoz2006,Gerard_2023}. Wavefront correction is provided by both segmented and continuous-face-sheet DMs, with the high-order DM (HODM) providing picometer-scale actuation. The testbed also incorporates both FAST and vortex coronagraphs for starlight suppression \cite{Gerard_2018}.

Beyond its hardware capabilities, HCT provides a platform for directly comparing multiple wavefront sensing architectures under identical optical and environmental conditions. This enables systematic evaluation of candidate WFSs for future ground- and space-based AO while leveraging the same DM hardware and coronagraphic optical train. These capabilities make HCT a powerful testbed for investigating the feasibility of picometer-level AOs relevant to future space-based observatories. 

\begin{figure}[ht]
    \centering
    \begin{subfigure}[b]{0.3\textwidth}
        \includegraphics[width=\textwidth]{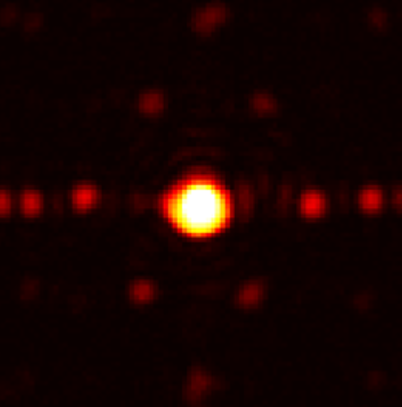}
    \end{subfigure}
    \hfill
    \centering
    \begin{subfigure}[b]{0.3\textwidth}
        \includegraphics[width=\textwidth]{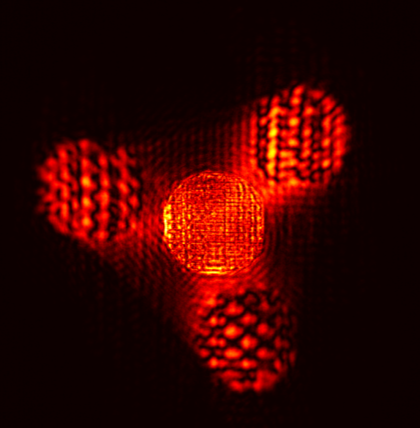}
    \end{subfigure}
    \hfill
    \begin{subfigure}[b]{0.3\textwidth}
        \includegraphics[width=\textwidth]{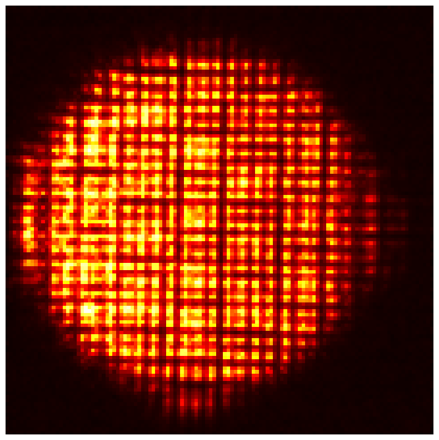}
    \end{subfigure}
    \vspace{6pt}
    \caption{Hight Contrast Testbed‘s (HCT) subset of wavefront sensors. (Left) PSF used for tip/tilt sensing for initial HCT stability. (Center) Pupil images formed by the 3RPWFS. (Right) Spot pattern produced by the SHWFS. }
\end{figure}

\subsection{Methodology}
The DMs were initially flattened using factory calibration files and optimized through focal-plane sharpening. A stabilized Helium-Neon laser served as the illumination source, producing a diffraction-limited point spread function (PSF). Following optimization, the Airy pattern and faint diffraction features associated with the DM actuator print-through were clearly visible, confirming diffraction-limited performance.

Although HCT is operated in air rather than under vacuum, the resulting low-frequency air turbulence induces temporal disturbances that are not representative of space-based instruments. Characterizing these disturbances establishes the current stability limit of the testbed while providing a benchmark for future environmental improvements. Temporal tip/tilt disturbances were then estimated directly from focal-plane images. Two independent measurement techniques were investigated: centroiding and phase cross correlation (PCC). Comparing these methods allows us to confirm their measurement sensitivities, identify potential testbed disturbances, and evaluate their measurement limitations. 

\subsubsection{Centroiding}

Centroiding estimates image translation from the first moment of the PSF intensity distribution. A region of interest (ROI) surrounding the PSF is used to reduce detector noise by excluding pixels containing little signal. Consequently, decreasing the ROI is expected to lower the measurement noise floor. Experimentally, the ROI is varied to determine the optimal trade-off between detector noise and measurement SNR. An excessively small ROI truncates portions of the diffraction pattern, potentially altering the centroid gain and reducing sensitivity to true tip/tilt motion, whereas an excessively large ROI incorporates unnecessary detector noise.

\subsubsection{Phase Cross Correlation}
Unlike centroiding, which estimates image motion using only the first spatial moment of the PSF intensity distribution, PCC estimates image translations by utilizing the Fourier-domain correlation between each measured PSF and a reference image. Because tip and tilt primarily produce lateral translations of an otherwise unchanged PSF, PCC naturally exploits the complete diffraction pattern—including the Airy core, diffraction rings, and higher-order intensity structure -- making it less sensitive to ROI selection and more representative of the underlying physical disturbances. 

\section{Results}

\begin{figure*}[ht]
    \centering
    \includegraphics[width=\columnwidth]{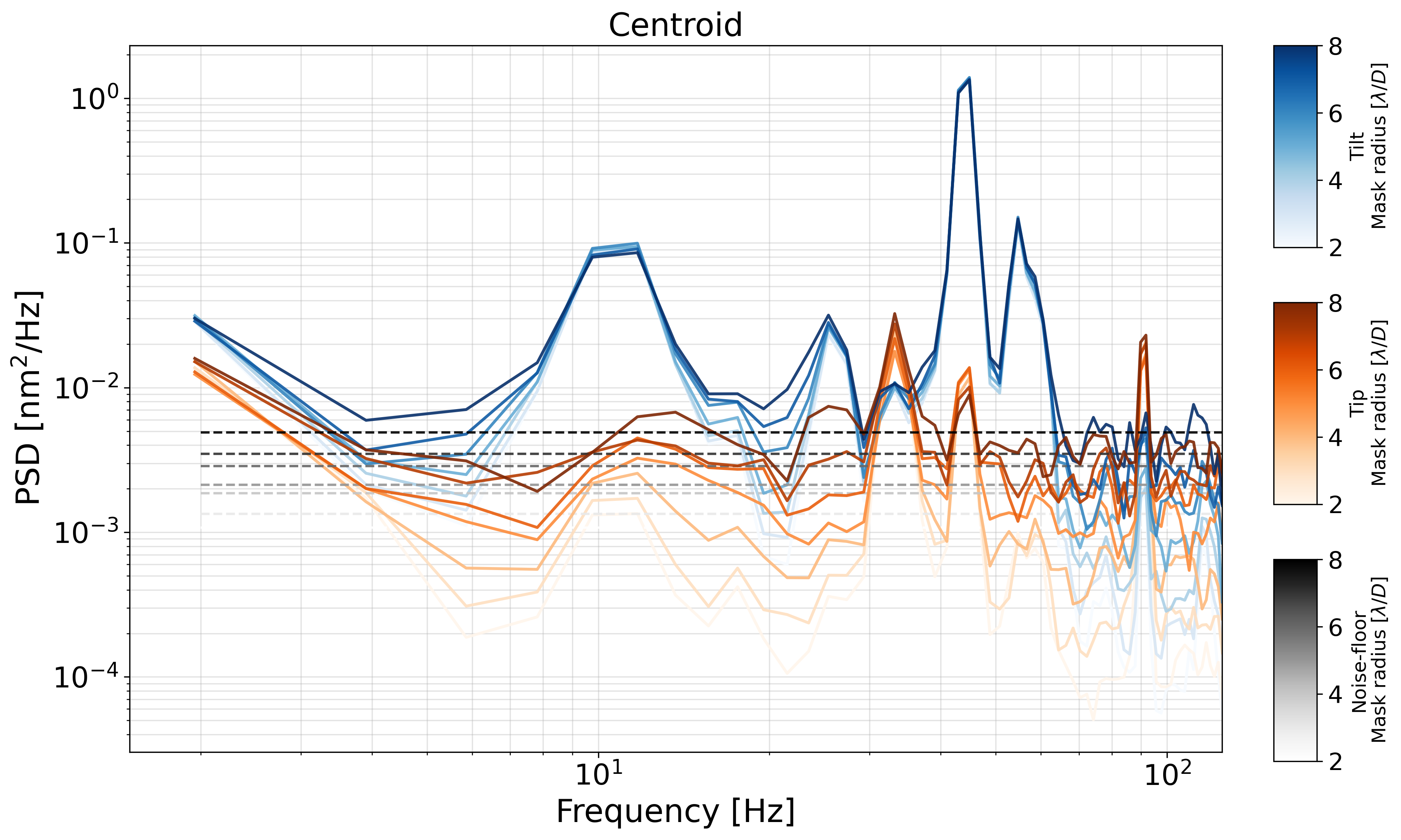}

    \vspace{0.5cm}

    \includegraphics[width=\columnwidth]{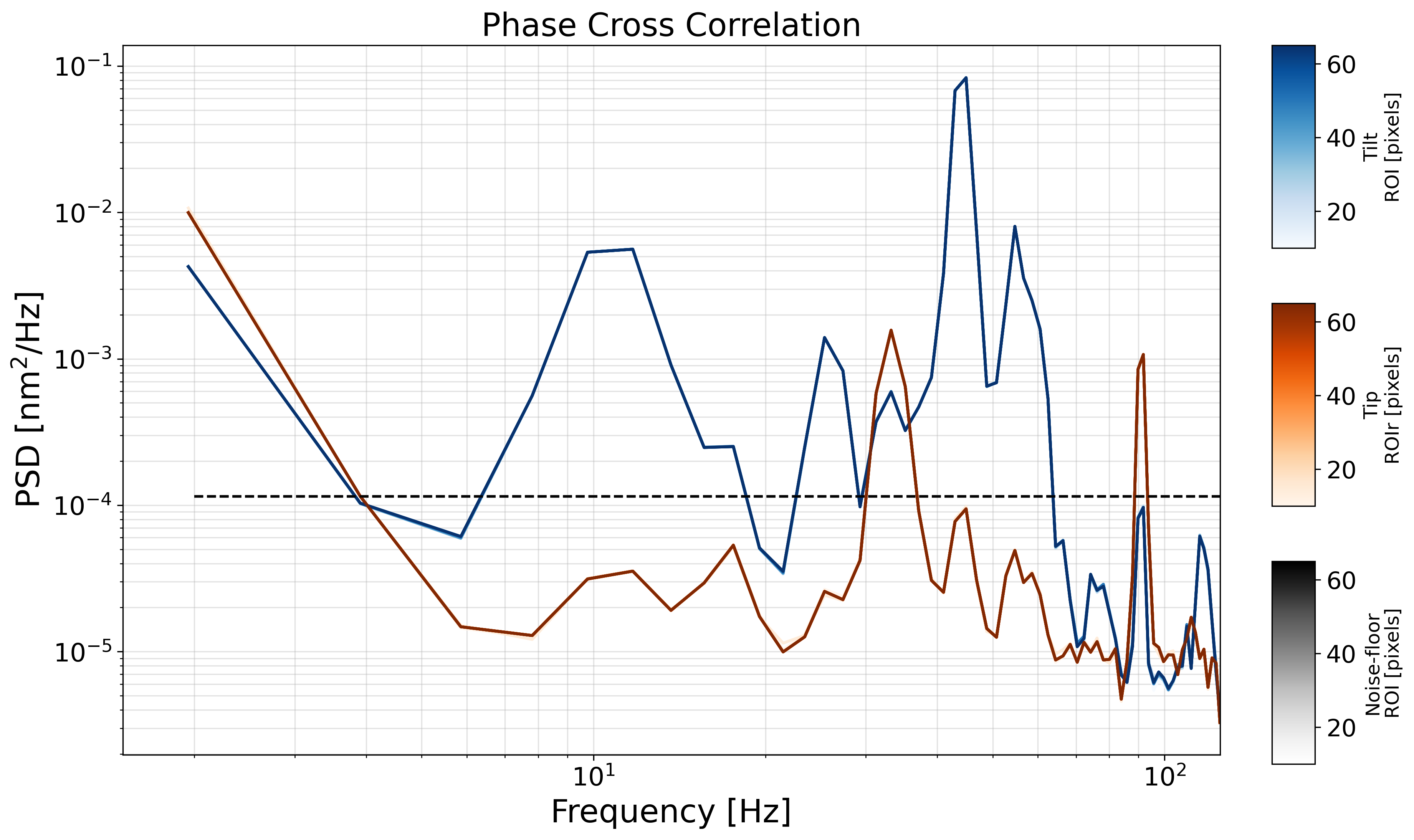}

    \caption{Comparison of temporal tip/tilt PSDs measured using (top) weighted centroiding and (bottom) PCC. For centroiding, reducing the ROI systematically suppresses the measured low-frequency disturbances while lowering the apparent noise floor, indicating that the estimator is sensitive to PSF truncation. In contrast, PCC preserves the physical disturbance spectrum across the range of ROIs investigated, suggesting that it provides a more accurate estimate of the temporal tip/tilt dynamics within the testbed.}
    \label{fig:stacked-psd}
\end{figure*}

Temporal power spectral densities (PSDs) were produced using tip/tilt measurements to characterize the testbed stability. At low frequencies, the spectrum is dominated by environmental disturbances, most likely caused by air turbulence within the enclosure. At sufficiently high frequencies, the spectrum reaches the detector noise floor, where the measured signal is no longer dominated by physical motion. The transition between these two regimes identifies the approximate disturbance bandwidth and a 10x larger AO control bandwidth is modeled to estimate the integrated error after correction -- assuming these disturbances have been suppressed to the noise floor. Figure~\ref{fig:stacked-psd} compares the temporal tip/tilt PSDs obtained using centroiding and PCC for a range of ROI sizes. The dashed horizontal lines indicate the estimated noise floor for each ROI, illustrating the influence of ROI size on both disturbance estimation and measurement sensitivity.

For centroiding, decreasing the ROI lowers the measurement noise floor as expected. However, the low-frequency portion of the PSD is also attenuated, suggesting that reducing the ROI suppresses the measured disturbance itself rather than solely improving measurement sensitivity. This behavior is consistent with finite-ROI truncation of the PSF, which effectively changes the centroid response and reduces the measured tip/tilt amplitude.

In contrast, the PCC measurements exhibit relatively little dependence on ROI size. The low-frequency disturbance spectrum remains nearly unchanged across the range of ROIs investigated, while the high-frequency noise floor varies only modestly. Unlike centroiding, no systematic attenuation or amplification of the physical disturbance is observed. This behavior indicates that PCC is less sensitive to ROI selection because it estimates image translations using the diffraction structure of the PSF rather than only its first spatial moment.

Both measurement techniques exhibit a transition from disturbance-dominated behavior to the measurement noise floor near 6~Hz. A representative 0~dB control bandwidth of 6~Hz and an AO frame rate of 60~Hz was therefore assumed for this initial analysis. Integrating the measured PSD to corresponding 30~Hz Nyquiest limit yields an estimated residual wavefront error of approximately 30-100~pm RMS, larger than the ultimate 10~pm stability goal. These measurements therefore establish the present temporal stability limit of HCT and identify the dominant environmental disturbances that must be mitigated before picometer-level AO can be demonstrated. 

In addition to low-frequency air disturbances, the measured PSD contains a distinct vibration near 45 Hz, particularly in the tilt axis. This feature is believed to originate from mechanical vibrations within the laboratory environment, although its source has not yet been identified. Because this narrow-band vibration is not representative of the measurement noise floor, it was excluded when estimating the high-frequency noise level.

\section{Conclusions}

This paper presented the current status of the mission concept, including preliminary mission architecture development, optical modeling of the laser guide star geometry, and initial laboratory demonstrations of temporal wavefront sensing.

The mission and optical analyses establish a first-order design space for the WaveDriver architecture. Preliminary mission studies illustrate the trade space between spacecraft count, servicing capability, and propulsion requirements. Optical modeling with EAC1 shows that LGS brightness, finite-conjugate effects, and field-angle constraints can be accommodated within practical operating regimes for the baseline HWO optical prescription. Together, these results indicate that the LGS concept is feasible with both the optical and mission-level requirements for future high-contrast space observatories.

Initial testbed measurements demonstrate progress toward the AO capabilities required for WaveDriver. Temporal PSD measurements shows the current stability limit of HCT while providing an initial comparison of focal-plane tip/tilt sensing algorithms, showing that PCC produces more robust disturbance estimates than centroiding. Future work will extend these efforts through higher fidelity mission and optical modeling, implementation of additional WFSs for higher-order measurement, and closed-loop demonstrations of picometer-level AO.

\acknowledgments 

This work was performed under the auspices of the U.S. Department of Energy by Lawrence Livermore National Laboratory under Contract DE-AC52-07NA27344. This document number is LLNL-PROC-2022488.   

\bibliography{spie2026_wavedriver_proceedings} 
\bibliographystyle{spiebib} 

\end{document}